\RequirePackage{ifpdf}
\documentclass{PoS}

\title{Precision measurements of cosmic ray air showers with the SKA}
\ShortTitle{Precision measurements of cosmic ray air showers}

\usepackage[authoryear,sort]{natbib}
\bibpunct{(}{)}{;}{a}{}{,}

\newcommand{\soton}{Univ.\ of Southampton}%{School of Physics \& Astronomy, Univ.\ of Southampton, SO17 1BJ, United Kingdom}
\newcommand{\nijmegen}{Radboud Univ.\ Nijmegen}%{Dept.\ of Astrophysics/IMAPP, Radboud Univ.\ Nijmegen, 6500 GL Nijmegen, The Netherlands}
\newcommand{\astron}{ASTRON}%{Netherlands Institute for Radio Astronomy (ASTRON), 7990 AA Dwingeloo, The Netherlands}
\newcommand{\karlsruhe}{KIT}%{IKP, Karlsruher Institut f\"ur Technologie, Postfach 3640, 76021 Karlsruhe, Germany}
\newcommand{\erlangen}{Univ.\ of Erlangen-Nuremberg}%{ECAP, Univ.\ of Erlangen-Nuremberg, 91058 Erlangen, Germany}
\newcommand{\groningen}{Univ.\ of Groningen}%{Kernfysisch Versneller Instituut, Univ.\ of Groningen, 9747 AA Groningen, The Netherlands}
\newcommand{\subatech}{Subatech, Nantes}%{Subatech, 4 rue Alfred Kastler, 44307 Nantes cedex 3, France}
\newcommand{\nancay}{Station de radioastronomie de Nan\c{c}ay}%{Station de radioastronomie de Nan\c cay, Observatoire de Paris, CNRS/INSU, Nan\c cay, France}
\newcommand{\atnf}{CSIRO ATNF}%{CSIRO Astronomy \& Space Science, NSW 2122, Australia}

\author{
 T.~Huege$^1$, J.D.~Bray$^2$, S.~Buitink$^3$, R.~Dallier$^{4,5}$, R.D.~Ekers$^6$, H.~Falcke$^{3,7}$, C.W.~James$^8$, L.~Martin$^{4,5}$, B.~Revenu$^4$, O.~Scholten$^9$ and F.G.~Schr\"oder$^1$\\
 $^1$\karlsruhe ;
 $^2$\soton ;
 $^3$\nijmegen ;
 $^4$\subatech ;
 $^5$\nancay ;
 $^6$\atnf ;
 $^7$\astron ;
 $^8$\erlangen ;
 $^9$\groningen \\
 E-mail: \email{tim.huege@kit.edu}%, \email{j.bray@soton.ac.uk}, \email{s.buitink@astro.ru.nl}, \email{richard.dallier@subatech.in2p3.fr}, \email{ron.ekers@csiro.au}, \email{h.falcke@astro.ru.nl}, \email{clancy.james@physik.uni-erlangen.de}, \email{lilian.martin@subatech.in2p3.fr}, \email{benoit.revenu@subatech.in2p3.fr}, \email{o.scholten@rug.nl}, \email{frank.schroeder@kit.edu}
}

\abstract{
Supplemented with suitable buffering techniques, the low-frequency part of the SKA can be used as an ultra-precise detector for cosmic-ray air showers at very high energies. This would enable a wealth of scientific applications: the physics of the transition from Galactic to extragalactic cosmic rays could be probed with very high precision mass measurements, hadronic interactions could be studied up to energies well beyond the reach of man-made particle accelerators, air shower tomography could be performed with very high spatial resolution exploiting the large instantaneous bandwith and very uniform instantaneous $u$-$v$ coverage of SKA1-LOW, and the physics of thunderstorms and possible connections between cosmic rays and lightning initiation could be studied in unprecedented levels of detail. In this article, we describe the potential of the SKA as an air shower radio detector from the perspective of existing radio detection efforts and discuss the associated technical requirements.

}

\FullConference{Advancing Astrophysics with the Square Kilometre Array,\\
		June 8-13, 2014\\
		Giardini Naxos, Sicily, Italy }

\begin{document}
% If the speaker is not the first author, then enter the first author
% information below, so it appears in the header of each page
% this must go AFTER the \begin{document} command
\makeatletter
\setbox\@firstaubox\hbox{\small T.~Huege}
\makeatother

\section{Introduction}

One hundred years after the discovery of cosmic radiation, the sources of the highest energy
particles are still a mystery. Previously observed correlations of cosmic-ray arrival directions
with catalogued positions of active galactic nuclei
at energies beyond $5 \cdot 10^{19}$~eV \citep{AugerCorrelation} have by now diminished in significance \citep{AugerCorrelationUpdate}.
At the same time, there are indications for a mixed proton-iron composition
up to the highest energies \citep{AugerComposition}, but better detectors will be needed to test this scenario.
Also, the transition from Galactic to extragalactic cosmic rays,
expected at energies between $10^{17}$ and $10^{19}$~eV, is not yet understood \citep{TransitionRegion}, in spite of decades of intense
research.

One reason for these many unsolved questions is the fact that there are a great many unknown
factors acting together: the physics of the acceleration at the sources, the mass composition of the
particle distributions from which cosmic rays are accelerated, the propagation through the little-known extragalactic and Galactic magnetic
fields, and in fact even the mass composition of the high-energy particles arriving at Earth. The last problem is due to
the fact that cosmic rays at very high energies arrive with an extremely low flux. They
can thus only be detected with ground-based detectors which do not measure the actual primary particle,
but an extensive air shower initiated by it. Determining the mass of the primary cosmic ray
from such indirect measurements, prone to uncertainties of hadronic interactions at energies
beyond the reach of man-made accelerators, has proven to be very difficult. Those detection techniques that
do provide a good determination of the primary particle mass, in particular detection of optical
fluorescence light, suffer from low duty cycles, limiting the badly needed statistics.

\begin{figure}[htb]
\centering
\includegraphics[clip=true,trim=0 0 0 380, width=0.7\textwidth]{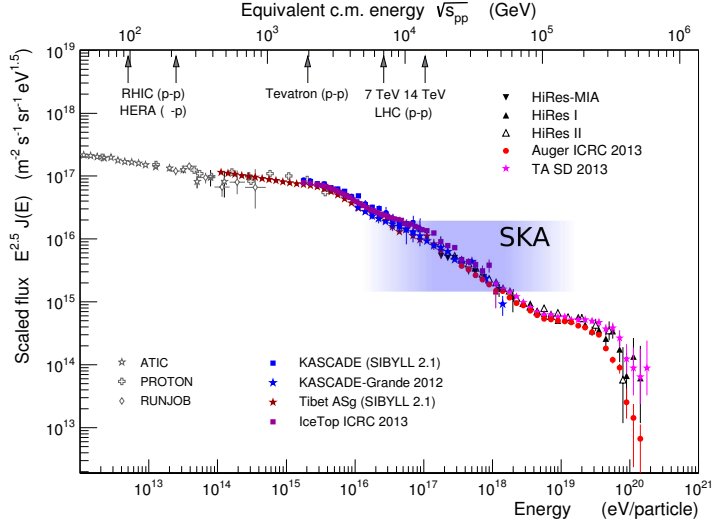}
\caption{The energy spectrum of the highest energy cosmic rays as updated and adapted from \citet{EngelSpectrum}. The energy range accessible to SKA1-LOW is indicated. This is the region in which the transition from Galactic to extragalactic cosmic rays is expected. Mass-composition-sensitive measurements with the SKA could determine the details of the transition with unprecedented accuracy. In addition, hadronic interactions can be studied up to energies much higher than those reachable with the Large Hadron Collider (LHC).}\label{fig:spectrum}
\end{figure}

Radio detection of cosmic rays has the potential to provide high-precision mass-sensitive
measurements with nearly 100\% duty cycle, as has recently been demonstrated with LOFAR \citep{lofar}. The low-frequency part of the
SKA will allow us to push these measurements to a new level, measuring cosmic-ray masses and studying hadronic
interaction physics with unprecedented precision up to energies as high as $10^{19}$~eV, cf.\ Fig.\ \ref{fig:spectrum}.
However, a number of technical requirements need to be fulfilled to use the SKA for cosmic-ray air shower detection.
In the following, we briefly lay out the current state of the field of radio detection of cosmic rays,
then describe the science goals for air shower detection with the SKA and afterwards describe the
technical requirements that would need to be fulfilled to reach these science goals.

\section{The state of cosmic ray radio detection}

Driven by the prospect of a precise and cost-effective detection technique for cosmic-ray air showers which measures with a duty cycle of nearly 100\%, digital radio detection of cosmic rays has matured from small prototype installations to experiments spanning several km$^2$ in the past decade \citep{huegeicrcreview}. Experiments such as LOPES \citep{falcke2005} and CODALEMA \citep{torres2013} have delivered the proof of principle for radio detection of cosmic rays, and have established the basic properties of the radio signal. The emission can be coherent up to GHz frequencies for specific geometries and is generally coherent in the frequency range up to $\sim 100$~MHz. It is dominated by radiation associated with the time-variation of geomagnetically induced transverse currents. In parallel with the experimental activities, sophisticated Monte Carlo simulation codes \citep{coreas, zhaires, selfas} as well as macroscopic models \citep{eva} have been developed to study the radio emission physics. Today, we have a very good understanding of the emission mechanisms, down to sub-dominant contributions such as the time-varying charge-excess radiation \citep[Askaryan effect;][]{askaryan}. Building on this understanding, it has been demonstrated that all important properties of extensive air showers can be deduced from radio measurements, in particular the arrival direction, energy, and mass-sensitive observables for individual cosmic rays \citep{lopes_ldf}.

\begin{figure}
\centering
\includegraphics[clip=true,trim=80 28 90 35, width=0.32\textwidth]{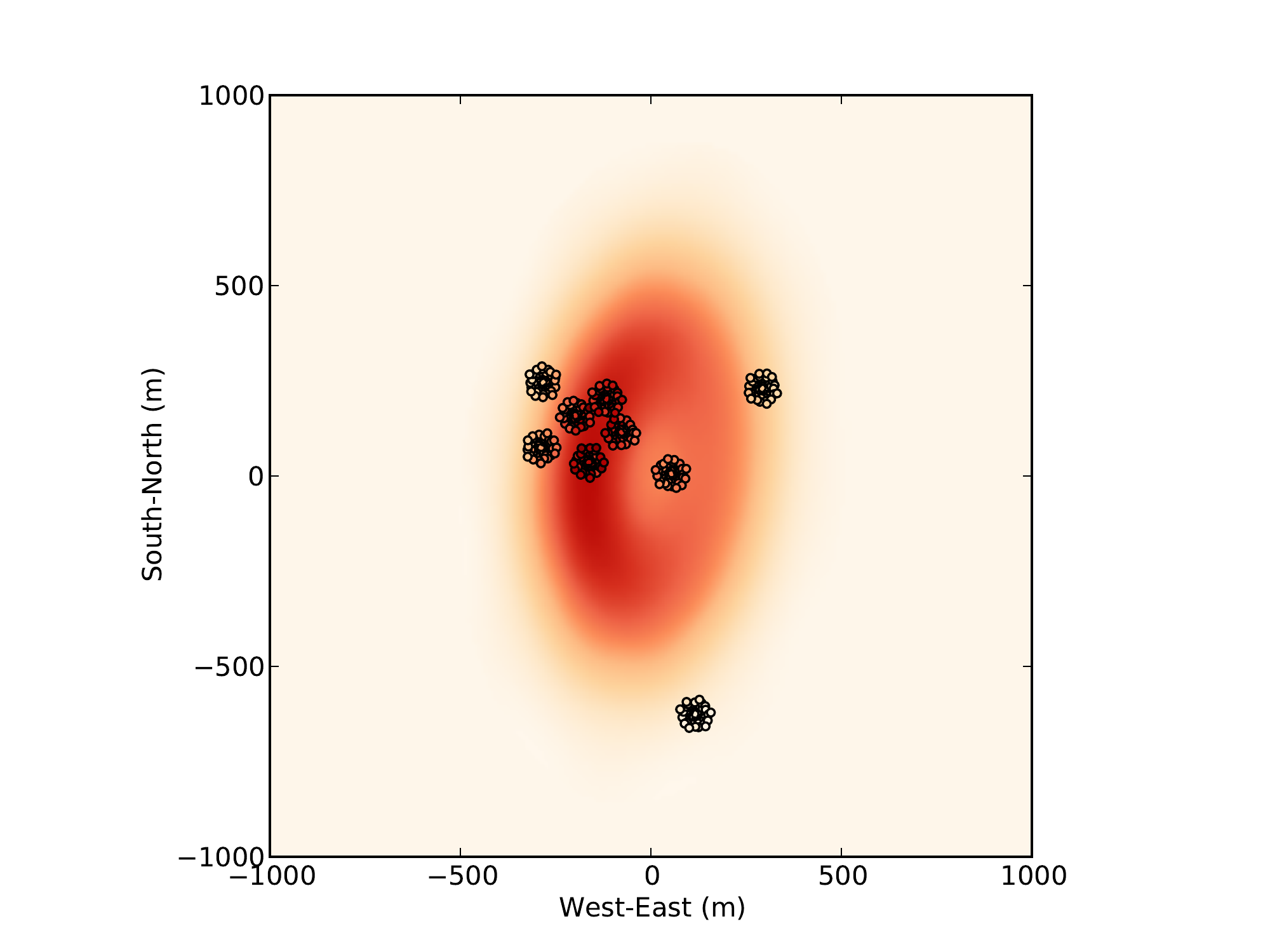}
\includegraphics[clip=true,trim=80 28 90 35, width=0.32\textwidth]{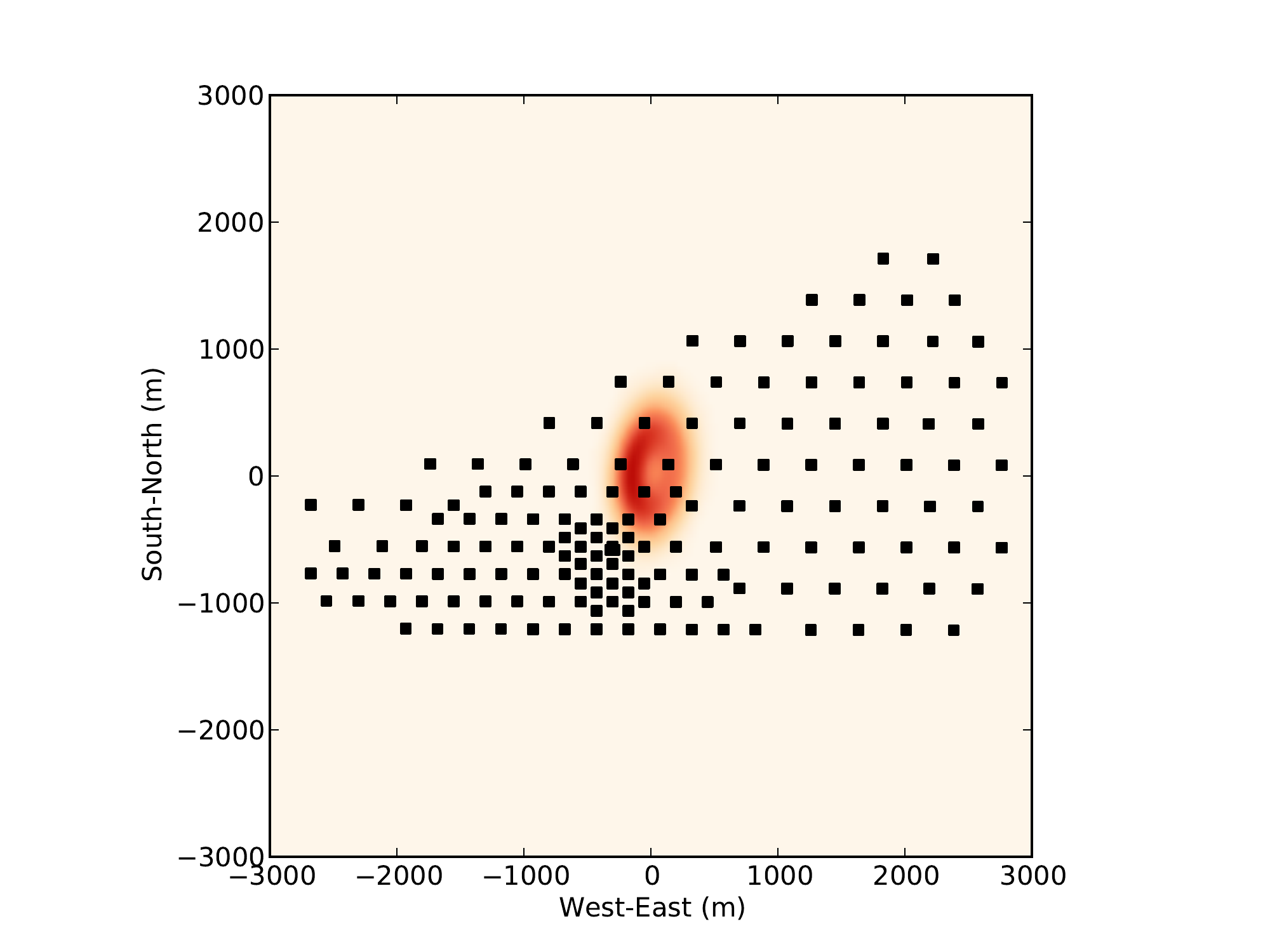}
\includegraphics[clip=true,trim=80 28 90 35, width=0.32\textwidth]{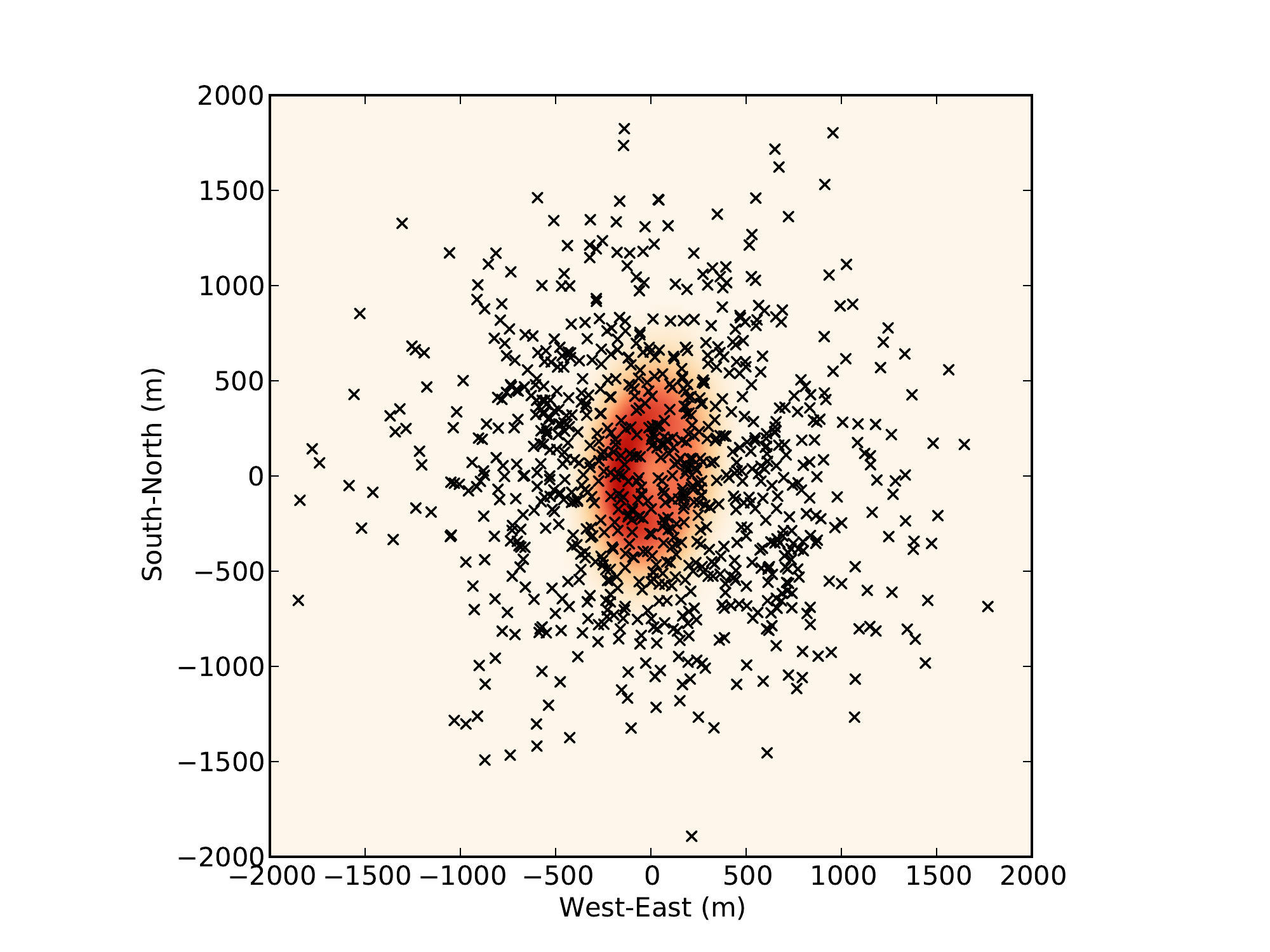}
\caption{Antenna layouts for several observatories (axes denote distances in metres). The antennas of LOFAR (left) are clustered in stations of 48 antennas each. Six of these stations lie in a dense core, additional stations are positioned at increasing distances. The background colors represent the radio footprint of an air shower simulated with CoREAS. The shape is elongated because the shower has a zenith angle of 55$^\circ$. The same footprint is shown in comparison with the AERA layout (middle), and the layout of SKA1-LOW (right), approximated by a Gaussian distribution of 866 antennas (i.e.\ a very small fraction of the actual number of antennas in the dense core). Both LOFAR and SKA sample the footprint with hundreds of antennas simultaneously, but at SKA the coverage is much more uniform. The instrumented area of SKA1-LOW is slightly smaller than AERA.}
\label{fig:detector_layouts}
\end{figure}

The major players in the current generation of experiments are the Auger Engineering Radio Array \citep[AERA;][]{aera} and LOFAR. AERA is focused at scaling the detection technique to much larger areas and thus energies and has by now instrumented an area of $\sim 6$~km$^2$ with a sparse array of autonomous antenna stations. One of its goals is to cross-check the mass sensitivity of the radio measurements directly with the Auger Fluoresence Detectors. In contrast, LOFAR features a very dense core of radio antennas over an area of $\sim 0.2$~km$^2$ allowing us to study the radio emission of individual air showers with much higher precision than any other experiment. It has recently been demonstrated that with these high-precision measurements, LOFAR is able to gather very high precision information on the mass of individual cosmic rays \citep{buitink_icrc}. A comparison of the scales and antenna densities of AERA, LOFAR and SKA1-LOW is shown in Fig.\ \ref{fig:detector_layouts}. With its very dense, uniformly instrumented core and its large instantaneous bandwidth, SKA1-LOW will be able to study individual air showers even more precisely than LOFAR, opening up the potential to address a number of long-standing questions in cosmic ray physics.

\subsection{LSS/NenuFAR as a pathfinder}\label{sec:nenufar}

The project ``NenuFAR'' --- the LOFAR Super Station (LSS) in Nan\c cay \citep{zarka2012} --- will be installed at the Nan\c cay Radio Observatory, which hosts a regular international LOFAR station (FR606), completely surrounded by the 57 radio detection stations of CODALEMA as well as 43 particle detectors. This environment provides a unique opportunity to test and improve techniques for cosmic ray radio detection, acting as a pathfinder for air shower detection with the SKA.

LSS/NenuFAR consists of an extension to the standard LOFAR local station including 96 mini-arrays of 19 crossed-dipole antennas (hereafter MA$_{19}$), analogically phased in the $\sim{10-87}$ MHz frequency window. These will be connected to the 96 dual-polarization receivers of the LOFAR backend. The 3648 dipoles will be positioned within a circle of 400 m diameter ($\sim{5}$ LOFAR station diameters). NenuFAR will be used both as part of the LOFAR network in LSS, and as a standalone instrument with large instantaneous sensitivity ($\sim{2}\times$ LOFAR's core) and wide frequency band. Standalone measurements will be possible due to a dedicated receiver, allowing NenuFAR to be used \textit{simultaneously} in both modes, within the instantaneous field of view of the beam formed at the MA$_{19}$ scale (10 to 50$^\circ$ from 10 to 87 MHz). As beam-forming is disfavoured for cosmic ray detection (see section \ref{sec:technical}), a dedicated acquisition channel will be implemented to extract the signal from one single antenna of each of the 96 MA$_{19}$ before beam-forming, and an external trigger input will be available to trigger NenuFAR by CODALEMA during regular astronomical operation. 
Beside this first cosmic ray detector operating mode, during dedicated standalone observations, NenuFAR can also be used in a ``radio fly's eye'' mode by phasing each of the 96 MA$_{19}$ in a different direction of the sky, thus covering the 2$\pi$~sr with 96 beams. The sensitivity will be 19 times higher than that reached with single antennas. The large field of view of each MA$_{19}$ ensures an overlap between adjacent beams and thus detection of distant showers by several MA$_{19}$. 

\section{Science goals for air shower detection with SKA1-LOW}

\subsection{Ultra-high resolution mass composition}

The SKA will have a densely populated core of $>1$~km$^2$, which makes it an excellent observatory for air showers in the energy regime of 10$^{16}$~eV to 10$^{19}$~eV. The detection rate will be $\sim 25,000$ showers above $10^{17}$~eV per year. Because of the steep drop-off of the cosmic-ray flux as a function of energy, there will be several hundreds of showers per year above 10$^{18}$~eV.
Below 10$^{17}$~eV, the radio detection efficiency of current-generation experiments is small because the signal strength becomes weak compared to the Galactic background noise. At the SKA, however, the sensitivity at low energies can be improved by beam-forming techniques and by exploiting the wide instantaneous bandwidth. Consequently, the detection threshold can likely be lowered well below $10^{17}$~eV.

The energy range of 10$^{17}$--10$^{19}$~eV probably harbors the transition from a Galactic to an extragalactic cosmic-ray component. While some models interpret the ankle feature in the all-particle spectrum at 10$^{18.2}$~eV as the transition energy, other models predict that the extragalactic component already becomes dominant at lower energies. In the latter class of models the ankle is caused by pair production of cosmic rays on the cosmic microwave and infrared background radiation \citep{ber88}. Furthermore, there may be a secondary Galactic component due to very energetic sources or re-acceleration of cosmic rays in the Galactic halo. To investigate how and at what energy the transitions between components take place, accurate mass composition measurements are crucial.

The main observable to study the mass of the primary particles initiating extensive air showers is the atmospheric depth (in g/cm$^2$) at which an air shower has evolved to its maximum particle number, the ``shower maximum'' $X_\textrm{max}$. LOFAR has demonstrated that with a dense antenna array, $X_\textrm{max}$ can be measured with a precision below 20~g/cm$^2$ by fitting simulated two-dimensional radio power profiles to the data \citep{buitink_icrc}. For showers with a particularly advantageous geometry, i.e.\ when the radiation profile on the ground is sufficiently well-sampled by antennas, the precision can be as good as 8~g/cm$^2$. More advanced analysis techniques that incorporate the polarization, spectrum and arrival time of the radio pulse will improve the reconstruction quality further. We expect that the application of such advanced techniques to the SKA will yield a mean precision of 10 g/cm$^2$. This is significantly better than the $X_\textrm{max}$ resolution achievable to date: The highest quality $X_\textrm{max}$ measurements are made with Fluorescence Detectors which achieve a resolution of $\sim 20$ g/cm$^2$, but with a duty cycle of only $\sim 10\%$. With the $X_\textrm{max}$ precision achievable with the SKA, it will be possible to measure the energy spectra of individual chemical elements or groups of elements with unprecedented precision. In particular, it will be possible to cleanly separate proton showers from other nuclei. This will allow the identification of the different source components between the second knee and the ankle.     

\subsection{Hadronic physics beyond the LHC scale}
The longitudinal development of air showers depends on several important features of hadronic interactions which are currently known only up to the energy that can be reached by the LHC ($\sim 10^{13}$~eV equivalent center-of-mass energy). Above this energy, several hadronic interaction models exist that provide extrapolated values based on phenomenological models, but the correct values cannot be calculated from first principles. The predictions that advanced models like QGSJET-II, EPOS and SYBILL make for shower properties such as $X_\textrm{max}$ and the muon-to-electron ratio diverge at super-LHC energies. \citet{ulrich2011} made a detailed study on how various shower properties are affected by the p-air cross section, the secondary particle multiplicity, the elasticity and the muon charge ratio. Accurate air shower measurements can thus probe hadronic physics at energies unreachable by man-made accelerators.    

A clean separation of protons from other cosmic ray particles (with a contamination at percent-level) allows measurement of the inclusive cross section for protons (p-air). At the Pierre Auger Observatory, the p-air cross section has been derived at a center-of-mass energy of 5.7~$\cdot 10^{13}$~eV, by measuring the tail of the $X_\textrm{max}$ distribution for protons \citep{AugerCC}. Their systematic uncertainty is dominated by the helium contamination of the shower sample. The SKA can measure $X_\textrm{max}$ with a higher resolution than Auger, allowing more precise measurements of the p-air cross section. 

In addition, it is expected that radio measurements can provide more information on the longitudinal development than only the depth of the shower maximum. The radiation is strongest when the number of charged particles in the shower changes most rapidly, at an early stage in the shower development. In general, the observed radio pulse contains information on all stages of the shower development. Near the shower core the signal is relativistically compressed but at larger distances the pulses are wider and the information is conserved. The information can be extracted by using a near-field imaging technique, for which the SKA is exceptionally well-suited due to its excellent instantaneous coverage of the $u$-$v$ plane (cf.\ Fig.~\ref{fig:detector_layouts}). Thus, ``air shower tomography'' can be used to extract more hadronic parameters. For example, the distance between the first interaction and the shower maximum depends critically on the secondary multiplicity and elasticity. Furthermore, shower imaging can be used to probe air-shower universality up to very high precision.

\subsection{Wide-band air shower measurements extending to high frequencies}

SKA1-LOW will have an instantaneous bandwidth of 50 to 350~MHz, which is much broader than current air-shower radio detectors. LOPES and AERA measure showers at frequencies below 100 MHz. LOFAR has both a low frequency (30--80~MHz) and high frequency window (110--250~MHz), but these cannot currently operate simultaneously. Recently, LOFAR has observed radio Cherenkov rings of $\sim100$~m radius in the high band \citep{LOFAR_HBA}. CROME \citep{smida2014} has detected GHz emission from air showers which was also found at typical distances of $\sim 100$ m to the shower core. The ANITA balloon experiment \citep{hoover2010} has picked up signals at frequencies from 300 to 900~MHz which are also likely to come from air showers; from their polarization mode it can be inferred that the signals do not originate from neutrino cascades inside the Antarctic ice, but are reflected off the ice surface and must originate from the atmosphere. 

Broad-band air shower measurements at the SKA can test air shower radiation models at frequencies up to 350~MHz with an unprecedented level of detail. Although various models are converging towards similar results at low frequencies, uncertainties remain at high frequencies. One reason for this is that the shorter wavelengths probe the shower on smaller scales and are thus more sensitive to small variations. In combination with the very uniform $u$-$v$ coverage, measurements up to 350~MHz will thus provide a tremendous advantage for the shower tomography technique described above. Another reason for an increased complexity of the radio emission at high frequencies is that additional emission mechanisms may start to contribute, like synchrotron radiation or molecular bremsstrahlung. A thorough understanding of the high-frequency emission will provide a calibration in particular for the events measured by ANITA.

\subsection{Cosmic rays and lightning initiation}

The severe thunderstorms that occur on the Australian mainland provide an additional research opportunity. It has been shown that atmospheric electric fields can influence the air-shower emission mechanism \citep{Buitink2007}. The strong fields present inside thunderclouds accelerate the shower electrons and positrons, and can strongly amplify the radio pulse. These effects also leave an imprint on the polarization, which can be used to infer the direction and possibly the strength of the atmospheric electric field. This provides a unique way to study the otherwise hard-to-measure electric fields in thunderclouds.

In addition, potential connections between air shower physics and lightning initiation can be studied with the SKA: air showers leave behind a trail of ionization electrons. Under normal weather conditions the electrons quickly attach to oxygen molecules, but inside electric fields they start to drift. If the field strength is high enough, the electrons will gain enough energy to ionize their environment further and create an electron avalanche. This process, known as runaway breakdown, may play an important role in the initiation of lightning \citep{RunawayBreakdown}. Radio arrays have the distinct advantage that they can measure both the air shower signal and the electrical processes in a thunderstorm.  With the SKA, the possible connection between air showers and lightning initiation can be explored with an unprecedented level of detail.  

Furthermore, the SKA can be used as a lightning mapping array which can make high-resolution 3D movies of the initiation of lightning and the electrical processes ultimately leading to a discharge. Because of the low attenuation of radio waves, the SKA can also search for high-altitude phenomena like sprites and discharges to the ionosphere.

%\begin{itemize}
%\item study possible connection between cosmic rays and lightning initiation
%\item study influence of atmospheric electric fields on air shower evolution and radio emission
%\end{itemize}

\section{Technical requirements}\label{sec:technical}

Detection of extensive air showers with SKA1-LOW has specific technical requirements which are significantly different from those of astronomical observations.
The signal emitted by cosmic-ray air showers is a very short transient with a length of ten nanoseconds up to a few hundred nanoseconds. As the emission source is in the atmosphere, there is no dispersion to longer time-scales as is the case for astronomical sources. It is also important to note that the area on the ground which is illuminated typically has a diameter of only 500 to 1000 metres. 

As the arrival direction of the cosmic-ray particles is unknown a priori, detection of air shower pulses in pre-beamformed data imaging only a limited field-of-view is strongly disfavoured. 
Instead, the best strategy is to read out the raw waveform information of individual antennas. These raw waveform data need to be read out with
high-resolution sampling of at least 700 MSPS and with a dynamic range of at least 8 effective bits above the noise level, i.e.\ ideally a raw dynamic range of 12 bits. The high dynamic range is needed because the amplitude of the radio signal emitted by air showers
varies very significantly over the illuminated area and in addition increases linearly with the primary particle energy. Thus, the waveform data has to be read out before any channelising or conversion to 2-bit dynamic range.

As also the arrival time is unknown a priori, it is necessary to buffer the individual antenna raw data for the time that is needed to wait for an external trigger to initiate read-out and
storage of the raw waveform data. This external trigger would need to be provided by an array of particle detectors that would need to be installed within SKA1-LOW. Particle detectors
are used routinely for cosmic ray shower detection and they do not generate any RFI, as has been successfully demonstrated at LOFAR. The time-scale on which the trigger
signal could be delivered, i.e.\ the required buffer depth, would be of order a few milliseconds.

To limit the memory needed for buffering, only a small fraction of the SKA1-LOW antennas would need to be buffered. Buffering antennas on a grid of order a few meters would be sufficient.\footnote{Consequently, even in the early deployment of SKA1-LOW with 50\% of the design sensitivity air shower detetection would be feasible to its full potential, provided that the instantaneous $u$-$v$ coverage of the buffered subset of antennas remains similar.} Once the trigger
signal arrives, it is also possible to only read out a subset of the buffers: using information about the location of the illuminated area (the so-called ``shower core''), only antennas in
an area with a radius of 500 to 1000 metres need to be read out and stored. To keep the buffer scheme flexible, the best strategy would be to feed all
individual antenna signals to a central location and provide a defined interface to access the unchannelised raw data of a user-defined subset of antennas for buffering purposes.

In summary, EAS detection with the SKA would require: access to high dynamic range raw waveform data of a user-defined subset of individual antennas, buffering capability for this subset of antennas for a few
milliseconds, a particle detector array for triggering purposes, and the ability to read out and store the raw waveform data of a subset of buffered antennas upon reception of a trigger signal. Suitable approaches will be tested on a reduced scale at NenuFAR, as mentioned in section \ref{sec:nenufar}.

\section{Conclusions}\label{sec:conclusion}

The SKA provides a unique opportunity to perform cosmic ray studies with very high precision using radio measurements in the 50-350~MHz band. In particular, ultra-high precision measurements of the mass composition of cosmic rays in the energy range from 10$^{17}$ to 10$^{19}$~eV would allow us to address the long-standing problem of the transition from Galactic to extragalactic cosmic rays. Furthermore, hadronic interactions could be studied up to energies well beyond the reach of the LHC, in particular if high-frequency air shower tomography using near-field imaging is exploited. Lastly, the physics of thunderstorms and the possible connections between cosmic ray physics and lightning initiation can be studied with the SKA. To enable these studies, millisecond-scale buffering of the high-sampling-rate, high dynamic range raw data of a subset of the individual SKA1-LOW antennas in the dense core is required.

%\newpage

\bibliographystyle{apj}

\end{document}